\documentclass{camera}
\usepackage{graphicx}  

\begin{document}

%
\title{Hybrid Detection of UHECR with the Pierre Auger Observatory}

%
\author{Miguel Mostaf\'{a} for the Pierre Auger Collaboration}

%
\organization{Physics Department, University of Utah\\Salt Lake City, UT 84112, USA}

\maketitle

\begin{abstract}
The Pierre Auger Observatory detects ultra-high energy cosmic rays
by implementing two complementary air-shower techniques.
The combination of a large ground array and fluorescence detectors,
known as the \textit{hybrid} concept,
means that a rich variety of measurements can be made on a single shower,
providing much improved information over what is possible with either detector alone.
In this paper the hybrid reconstruction approach and its performance are described.
\end{abstract}

%
\section{Introduction}

The Pierre Auger Observatory
was designed to observe, in coincidence, the shower particles at ground
and the associated fluorescence light generated in the atmosphere.
This is
achieved with a large array of water Cherenkov detectors coupled with
air-fluorescence detectors that overlook the surface array.
It is not simply a dual experiment.
Apart from important cross-checks and measurement redundancy,
the two techniques see air showers in complementary ways.
The ground array measures the lateral structure of the shower at ground level,
with some ability to separate the electromagnetic and muon components.
On the other hand, the fluorescence detector records the longitudinal pro{f}ile of the shower during
its development through the atmosphere.

\begin{figure}[!ht]
	\centerline{%
\includegraphics[width=0.4\textwidth]{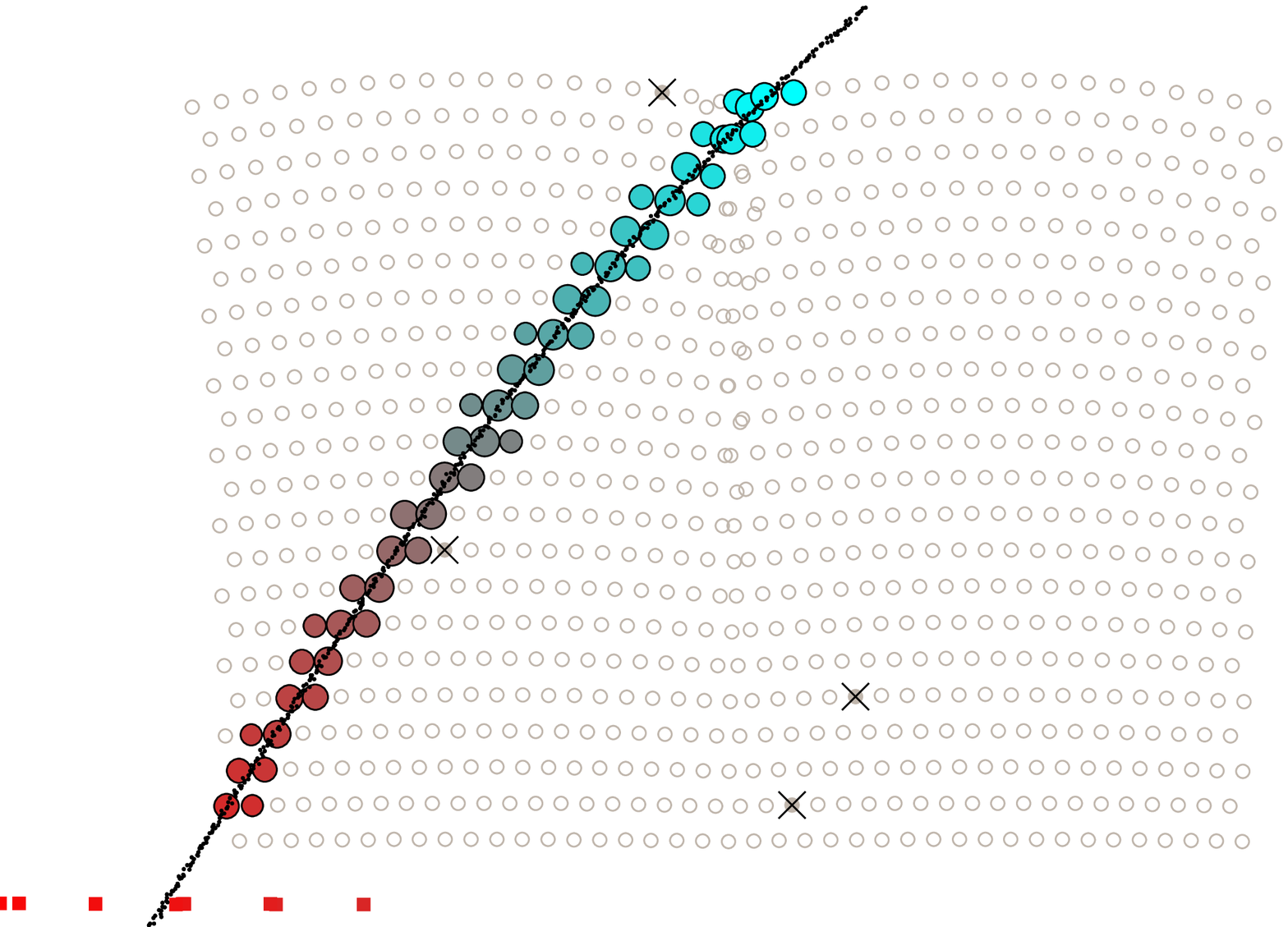}%
\includegraphics[width=0.4\textwidth]{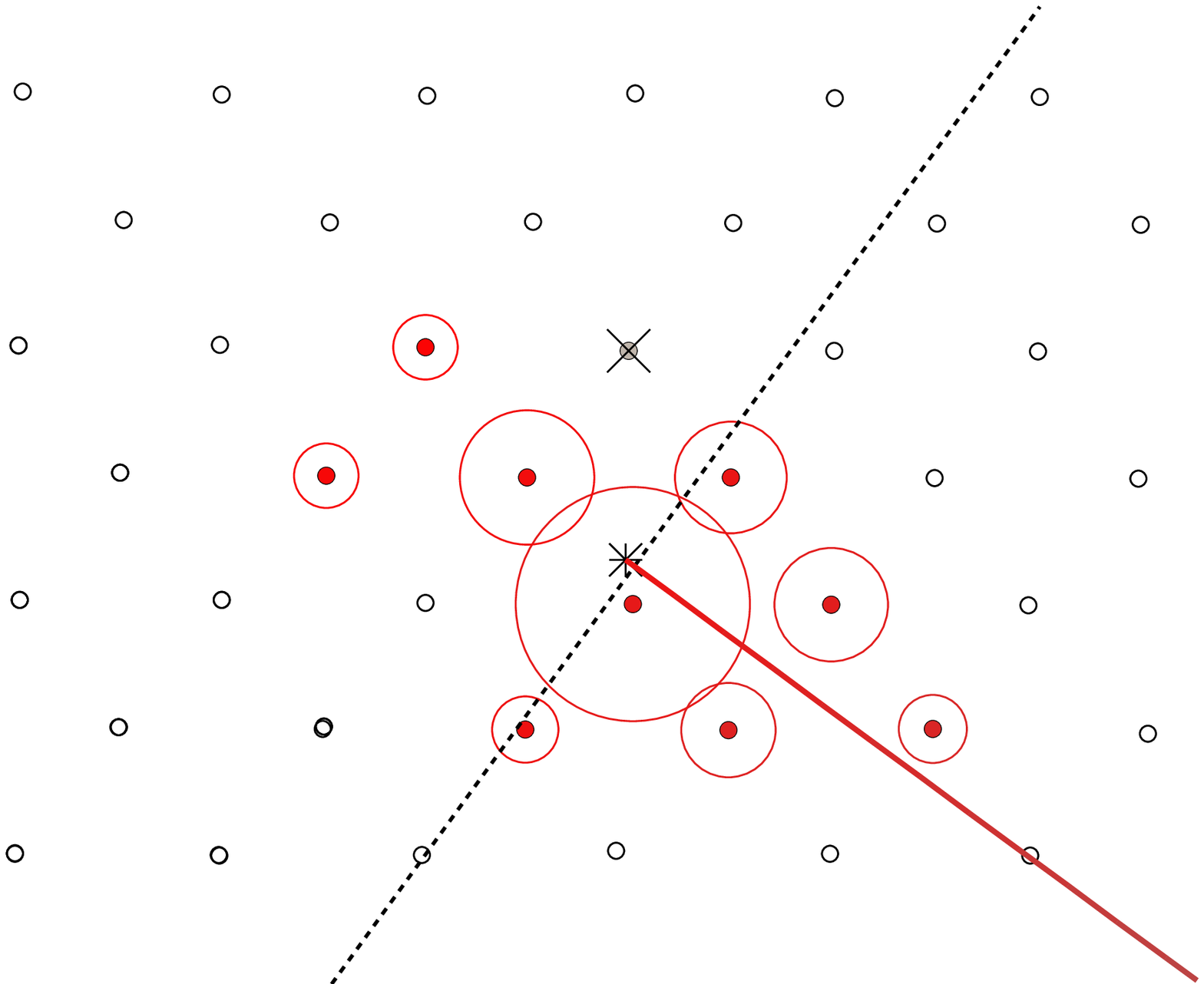}}
\caption{Example of an air-shower that independently triggers the fluorescence and surface detectors.
\textit{Left:} Fluorescence detector view.
The pixel color indicates the timing sequence.
The full line is the fitted shower-detector plane.
The red squares represent the surface detectors.
\textit{Right:} Surface detector view.
The size of the hollow circles is proportional to the logarithm of the signal in each tank.
The dotted line is the intersection between the shower-detector plane and the horizontal.
The full red line is the projection of the shower axis,
and the star marks the location of the shower core.}
\label{fig:golden} 
\end{figure}

A \textit{hybrid} event is an air shower that is simultaneously detected
by the fluorescence detector and the ground array.
The Observatory was originally designed and is currently being built with a \textit{cross--triggering}
capability.
Data are recovered from both detectors whenever either system is triggered.
If an air shower independently triggers both detectors the event is tagged
accordingly.
An example of this type of events,
known as \textit{golden} hybrids,
is shown in Fig.~\ref{fig:golden}.

\begin{figure}[!hb]
	\centerline{%
\includegraphics[width=0.4\textwidth]{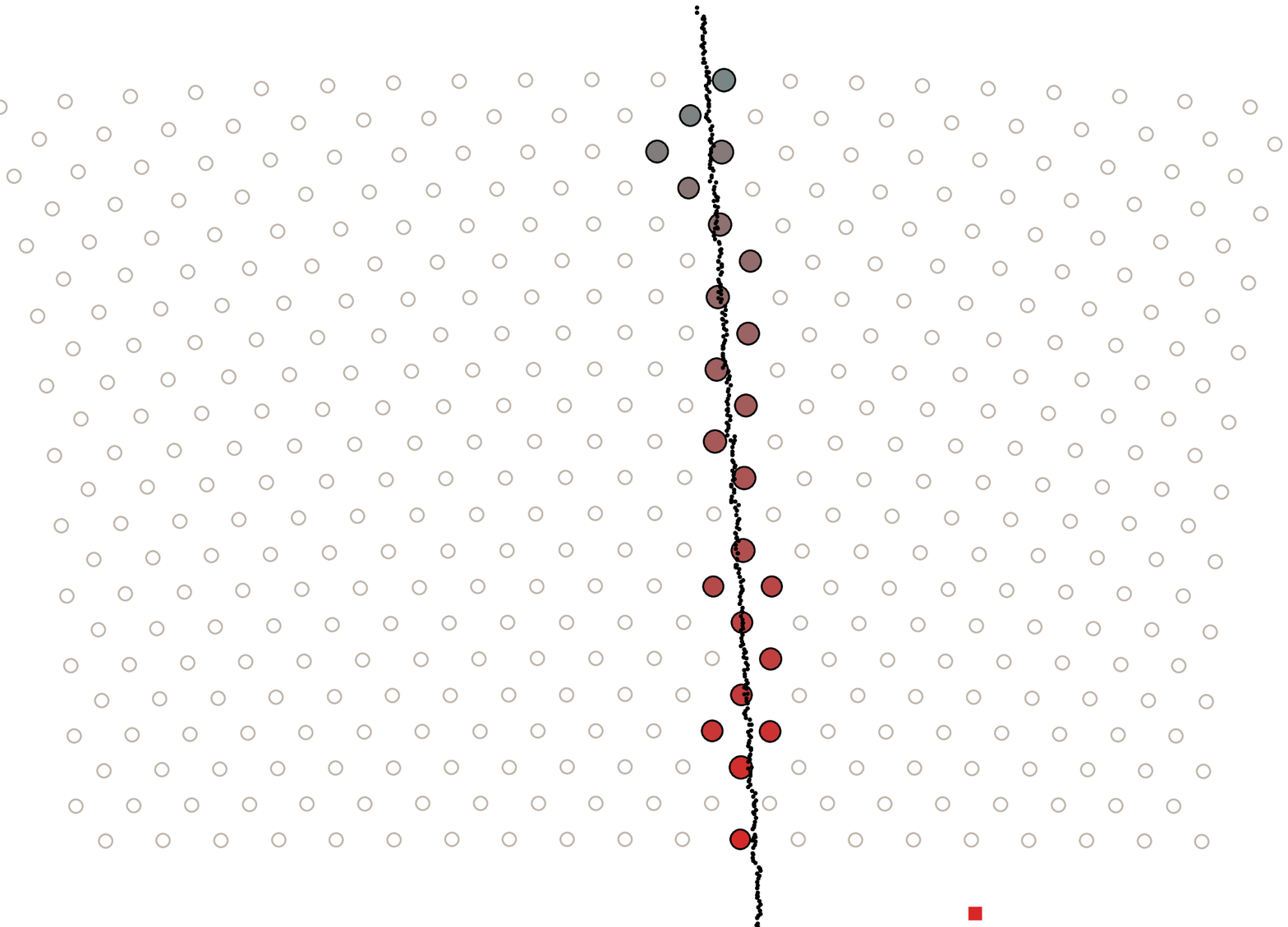}%
\includegraphics[width=0.4\textwidth]{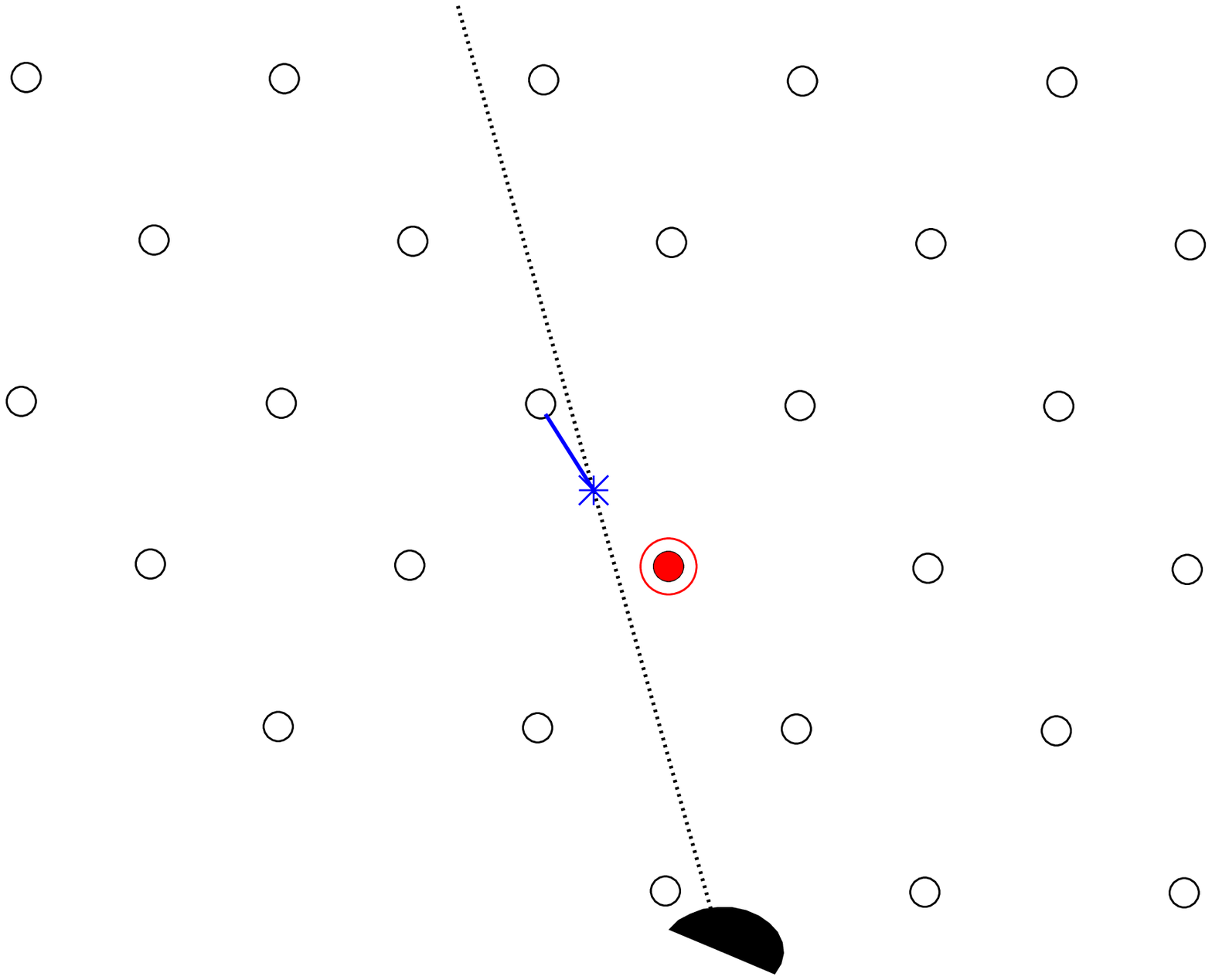}}
\caption{Example of a \textit{sub-threshold} hybrid event.
The trigger from the fluorescence detector is matched in time and location by only one of the water tanks.
In the array view (\textit{right}) the fluorescence detector is shown as a black semi-circle.}
\label{fig:subt} 
\end{figure}

There are also cases where the fluorescence detector,
having a lower energy threshold,
promotes a sub--threshold array trigger.
Surface stations are matched by timing and location.
This is an important capability because these sub--threshold hybrid events
would not have triggered the array otherwise.
One example is shown in Fig.~\ref{fig:subt}.

The Observatory started operation in \textit{hybrid} production mode in January, 2004.
Surface stations have a $100$\% duty cycle,
while fluorescence eyes can only operate on clear moonless nights.
Both surface and fluorescence detectors have been running simultaneously
$14\%$ of the time.
The number of hybrid events represents $10\%$ the statistics of the surface array data.

\section{Hybrid Detection}

A hybrid detector has excellent capability for studying the highest energy
cosmic ray air showers.
Much of its capability stems from the accurate geometrical reconstructions
it achieves.
Timing information from even one surface station can much improve the geometrical
reconstruction of a shower over that achieved using only eye pixel information.
The axis of the air shower is determined
by minimizing a $\chi^{2}$ function involving data from all triggered elements in the eye
and at ground.
The reconstruction accuracy
is better than the ground array counters
or the single eye could achieve independently~\cite{Bellido:2005aw,Bonifazi:2005ns}.
Using the timing information from the eye's pixels together with the surface stations,
a core location resolution of $50$~m is achieved.
The resolution for the arrival direction of cosmic rays is
$0.6^{\circ}$~\cite{Bonifazi:2005ns}.
These results for the \textit{hybrid} accuracy are in good agreement with
estimations using analytic arguments~\cite{Sommers:1995dm},
measurements on real data using a bootstrap method~\cite{Fick:2003qp},
and previous simulation studies~\cite{Dawson:1996ci}.

The reconstruction uncertainties are evaluated
using events with \textit{known} geometries, \textit{i.e.} laser beams.
The Central Laser Facility (CLF), described in Ref.~\cite{Arqueros:2005yn},
is located approximately equidistant from the first three fluorescence sites.
Since the location of the CLF and the direction of the laser beam are
known to an accuracy better than the expected angular resolution of the
fluorescence detector, laser shots from the CLF can be used to measure the accuracy of
the geometrical reconstruction.
Furthermore, the laser beam is split and part of the laser light is sent
through an optical fiber to a nearby ground array station.

The laser light from the CLF produces simultaneous triggers in both the surface and
(three) fluorescence detectors. The recorded event times can be used to measure and monitor the
relative timing between the two detectors.
The time offset between the first fluorescence eye and the surface detector 
has been measured in this way to
better than $50$~ns~\cite{Allison:2005vj}.
The contribution to the systematic uncertainty in the core location due
to the uncertainty in the time synchronization is $20$~m.

Hybrid events can be well reconstructed with as little as one surface station.
Thus,
the energy threshold of hybrid events is lower than
it is for surface array only events,
where at least three stations must be triggered.
Sub--threshold hybrid events have a lower energy distribution,
while the angular resolution is still $0.6^{\circ}$.
Another important consequence is the additional number of events.
Approximately $60\%$ of the total hybrid events have fewer than three stations,
\textit{i.e.} for each event that can independently trigger both
the surface array and one fluorescence detector there are two extra low energy
events.

\begin{figure}[!ht]
\centerline{\includegraphics[width=0.4\textwidth]{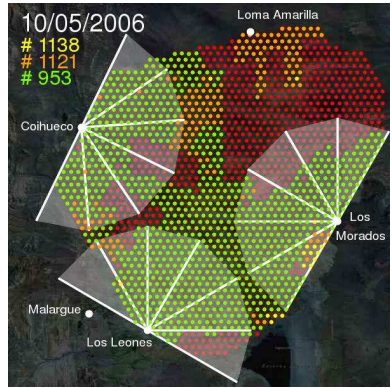}}
\caption{\label{STATUS} Status of the Pierre Auger Observatory on May 10, 2006.
Each color dot represents a surface station:
\textit{red} means a proposed location,
\textit{yellow} a deployed tank,
\textit{orange} a tank with water (but without electronics), and
\textit{green} a fully operational tank.
Each fluorescence site is shown as a white circle.
}
\end{figure}

The status of the Observatory on May 10, 2006 is summarized in Fig.~\ref{STATUS}.
There were 953 fully operational tanks at that time.
The {f}irst two fluorescence sites (Los Leones and Coihueco) were fully operational,
\textit{i.e.} running six telescopes each, in June, 2004.
The third site (Los Morados) started operation on March 18, 2005.
The fourth and last site (Loma Amarilla) is currently under construction and is scheduled
to start operation by the end of this year.
The present average rate is $50$ hybrid events per night per eye,
for a total of $\sim40000$ events up to June, 2006.
At this rate, $\sim4000$ hybrid events per month are expected
when the Observatory is completed.

\section{Hybrid Measurements}

Due to the much improved angular accuracy,
the \textit{hybrid} data sample is ideal for anisotropy studies
and, in particular, for point source searches.
Results on a search for a point--like source in the direction of the galactic
center using these hybrid events were presented in Ref.~\cite{Letessier-Selvon:2005uf}.

Many ground parameters, like the shower front curvature and thickness, have always been dif{f}icult to measure
experimentally, and were usually determined from Monte Carlo simulations.
The hybrid sample provides a unique opportunity in this respect.
As mentioned, the geometrical reconstruction can be done using only one ground station,
thus all the remaining detectors can be used to measure the shower characteristics.

The combination of the air fluorescence measurements and particle detections on the ground
provides an energy measurement almost independent of air shower simulations.
The fluorescence measurements determine the longitudinal development of the shower,
whose integral is proportional to the total energy of the electromagnetic particle cascade.
At the same time,
the particle density at any given distance from the core
can be evaluated
with the ground array.
The conversion from particle density to the energy of the shower
is where the fluorescence measurements become important.
Hybrid events that can be independently reconstructed with both techniques
are used to establish an empirical rule for the energy conversion.
The procedure to determine the energy of each event is explained in more detail
in Ref.~\cite{Sommers:2005vs}.
It is important to note that
both techniques have different systematics,
and results are preliminary at this stage while the Observatory is under construction.
The possibility of studying
the same set of air showers with two independent methods is
valuable in understanding the strengths and limitations of each technique.
The \textit{hybrid} analysis bene{f}its from the calorimetry of the fluorescence technique
and the uniformity of the surface detector aperture.

\section{Conclusions}

We have tested the performance of Pierre Auger Observatory
in its hybrid con{f}iguration.
Operation started in January, 2004 and over $40000$ hybrid events have been successfully reconstructed
up to now.
It is important to note that the Observatory is under construction
and that results are preliminary.
It is already clear that the combination
of fluorescence and ground array measurements provides reconstruction
of the geometry of the shower with much greater accuracy than is achieved
with either detector system on its own.
Unprecedented core location and direction precision
leads to excellent energy and shower development measurements.
Several Auger measurements 
pro{f}it from the hybrid capability of the Observatory,
including the
angular resolution~\cite{Bonifazi:2005ns},
energy spectrum~\cite{Sommers:2005vs}, and photon limits~\cite{Risse:2005hi}.

%
\end{document}